# Brillouin-enhanced four-wave mixing with optical chiral states

**XINGLIN ZENG**[1,*] **AND BIRGIT STILLER**[1,2,3]

[1]*Max-Planck Institute for the Science of Light, Staudtstr. 2, 91058 Erlangen, Germany*
[2]*Department of Physics, Friedrich-Alexander-Universität, Staudtstr. 7, 91058 Erlangen, Germany*
[3]*Institute of Photonics, Leibniz University Hannover, Welfengarten 1A, 30167 Hannover, Germany*
*xinglin.zeng@mpl.mpg.de

**Abstract:**

Brillouin-enhanced four-wave mixing - also known as Brillouin dynamic gratings - is an important nonlinear effect in photonics that couples four light waves by travelling acoustic waves. The effect has received a lot of attention in the last few decades, especially for applications in fiber sensing, signal processing and optical delay lines. Here, we report Brillouin-enhanced four-wave mixing with optical chiral states (i.e. circular polarization and vortex states) in twisted photonic crystal fiber, by leveraging the topology-selective Brillouin effect. Phase-matching has the consequence that the travelling acoustic gratings created by circularly-polarized vortex pump and Stokes in the stimulated Brillouin scattering can be used to modulate a frequency-shifted probe, where the pump/Stokes and probe have different circular polarization or topological charges. We demonstrate cross-frequency selective information transfer and show that the information is transferred only when pump and probe have opposite circular polarization.

## 1.  Introduction

Brillouin-enhanced four-wave mixing (BE-FWM) is a physical phenomenon resulting from stimulated Brillouin scattering (SBS) in an optical medium, where two counter-propagating light waves each at different frequency create a travelling acoustic grating that can act as a transient reflector or memory for subsequent optical signals, storing and manipulating information within the medium [1–3]. BE-FWM experiments were initially conducted in bulk materials and liquid cells to study optical phase conjugation properties [4] and beam combining effect [5]. It was not until the 2000s, when BE-FWM was demonstrated in optical fibers (also named Brillouin dynamic gratings) [6,7], which was followed by applications in optical storage [8], signal processing [9] and optical sensing [10]. The Brillouin dynamic grating (BDG) could be measured or detected by monitoring the diffracted wave from a third probe wave which is used to illuminate the grating. Therefore, most BE-FWM experiments are based on linearly-polarized fundamental modes in polarization maintaining fiber [7,11], such that the pump and probe waves can be easily distinguished due to the refractive index difference between two orthogonal linear polarization states. Some BDG experiments have also been demonstrated in few-mode fibers (FMF) as there are large refractive index differences between the different fiber modes [12].

Optical chiral states, such as the optical waves carrying spin angular momentum (SAM) or orbital angular momentum (OAM), are attracting extensive attentions over the last decade, in connection with applications in quantum and classical communications, optical nanomanipulation, and quantum information processing [13–15]. In recent years, chiral photonic crystal fiber (PCF) [16], drawn from a spinning preform, has emerged as a unique platform to study the behavior of light in chiral structures that are capable of preserving circularly-polarized vortex states carrying both SAM and OAM. Over the last years, a series of studies on SBS have been conducted in chiral PCFs, showing topology-selectivity and angular momentum transfer capability in the Brillouin coupling effect in the presence of chirality [17–20].

Here, we report the first experimental study of BE-FWM with optical chiral states by leveraging the circular polarization-selective and topology-selective SBS effects in twisted PCF. The travelling BDG generated by SBS between two identical and counterpropagating circular polarization states/vortex modes diffract a third probe signal and generate the fourth readout signal. Phase-matching requires that the probe signals have different frequencies and carry different circular polarization states or different vortex states. We demonstrate frequency-shifted information-selective transfer with the inter-circular-polarization BE-FWM effect and the data reading can be switched-on and off by tuning the polarization states of probe signal. The experimental results offer a new perspective for BE-FWM or BDG and are of potential interests in advanced information processing, optical sensing and quantum manipulation.

## 2. Results

N-fold rotationally symmetrical (symmetry class $C_N$) chiral PCFs are capable of supporting helical Bloch modes (HBMs) that carry circularly-polarized vortex states, which are chiral optical states with SAM and OAM [21]. We use [$\ell$, $s$] to denote the HBMs, where the $\ell$ is topological charge and $s$ is spin order. In twisted PCF, HBMs with equal and opposite values of $\ell$ are generally nondegenerate in index (i.e., topologically) while modes with opposite spin but the same value of $\ell$ are weakly birefringent. Fig. 1(a) show the schematic of BE-FWM with optical chiral states in the twisted PCF. We use the superscripts "+" and "−" to denote forward and backward propagation, rather than changing the signs of mode parameters directly as in previous papers [18]. In fact, both mode labelling methods are physically meaningful, but the former one is more appropriate for interpreting the results of this paper. The BE-FWM is an effect where four optical waves interact with each other through two SBS processes. Counter-propagating pump and Stokes with frequency difference $\Omega_B$ interact and generate a BDG through the first SBS process. Both pump and Stokes carry left circular polarization states. In the second SBS process, the generated BDG can back-act on a frequency shifted probe wave of orthogonal circular polarization states (frequency shift $\Delta\Omega_1$) or different topological charges (frequency shift $\Delta\Omega_2$) to pump/Stokes, while generating a fourth readout wave that is the same as the probe but propagates in opposite direction. The frequency difference between probe and readout is also $\Omega_B$ because the entire nonlinear process is mediated by only one acoustic wave. Fig. 1(b) shows the dispersion diagram illustrating the principle of phase-matched readout of BDG at different wavelength. The solid curves are dispersion relations of optical modes and the red dashed curve is the wavevector of BDG for phase matching. The acoustic wave generated by pump and Stokes can be used for frequency-shifted SBS between other chiral optical modes (probe and readout) that have different dispersion relations. The phase matching wavelength shift from probe to pump is $\Delta\lambda_i = (n_i-n_P)\lambda_P/n_P$, with the corresponding frequency shift $\Delta\Omega_i = (n_i-n_P)v_P/n_P$, where the $n_i$ is the refractive indices of different probe waves and $v_P$ is the pump frequency.

Fig. 2(a) shows the experimental setup for BE-FWM with chiral optical states. We used a three-fold rotationally symmetrical twisted ($C_3$) PCF, which has a central solid core and three satellite cores. The fiber has a length of 200 m, twist pitch of 5 mm, hollow channel diameter of 1.6 μm, and inter-channel spacing of 5.6 μm. The inset in the right-side of Fig. 2(a) shows the scanning electronic micrograph (SEM) of the fiber cross-section. The output from a continuous wave (CW) laser at 1550 nm was split into pump and Stokes signals. The Stokes was frequency down-shifted by $\Omega_B$ in a single sideband modulator (SSBM). Both signals were boosted by erbium-doped fiber amplifiers (EDFA) and then launched oppositely into twisted PCF to generate BDG through SBS effect. The fiber polarization controllers (FPC) were used to tune the circular polarization states. The vortex generation modules (VGM), which consist of a polarizer, a λ/4 plate and a Q-plate, are optionally used to generate circularly-polarized vortex beams. In the intervortex BE-FWM experiments, the probe wave is generated by a wavelength-tunable laser; in inter-circular-polarization BE-FWM experiments requiring small frequency shifts (<GHz) and fine- frequency-tuning, the probe wave is generated by a combination of a CW laser and an intensity modulator (IM). Two λ/4 plates and a polarizing beam splitter (PBS) were used to separate the readout wave from probe wave. After passing through a tunable filter, the readout wave was measured by a high-resolution optical spectrum analyzer (OSA) or power meter. Fig. 2(b) shows the spontaneous Brillouin spectra generated by pumping [0,+1], [−1,−1] and [+1,+1] modes at 1550 nm, with Brillouin frequency shifts of 11.066 GHz, 11.056GHz and 11.052 GHz. The circular birefringence between fundamental modes and vortex modes are ~3.5×10$^{-6}$ and ~5×10$^{-6}$. The refractive indices of all six modes are shown in the Appendix A. Due to the weak circular birefringence, the Brillouin spectra of the modes ([0,−1], [+1,−1] and [−1,−1]) are very close to the other three, not shown in the Fig. 1(b).

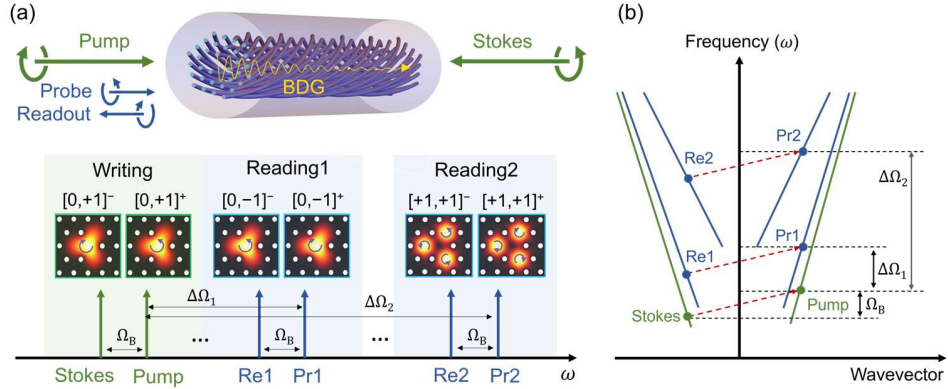

**Fig.** 1. (a) Basic schematic of BE-FWM with chiral optical states in twisted PCF. The insets in the color-shaded areas show the calculated mode profiles of writing and readout chiral modes; (b) Dispersion diagram illustrating the inter-circular polarization and inter-vortex BE-FWM.

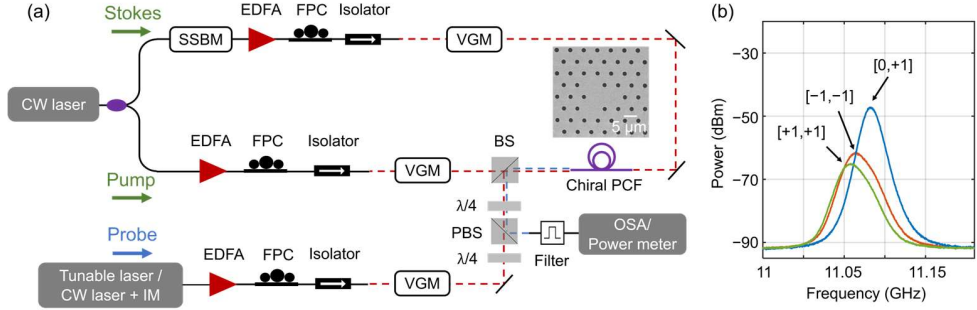

**Fig.** 2. (a) Experimental setup for BDG of chiral optical states (circular polarization states and vortex states). The inset is the scanning electronic micrograph image of the twisted PCF cross section. (b) Spontaneous Brillouin spectra of [0,+1], [−1,+1] and [+1,+1] measured by experimental setup in [18].

We first measured the inter-circular-polarization BE-FWM in twisted PCF by using an OSA. An intensity modulator (IM) was used to shift the frequency of probe wave from the pump. Fig. 3(a) shows the measured spectrum after filter when forward [0,−1]$^+$ pump, backward [0,−1]$^-$ Stokes and forward [0,+1]$^+$ probe are launched into the fiber. The backward [0,+1]$^-$ readout wave is generated. The subscripts "Pm", "St", "Pr" and "Re" in the figure denote pump, Stokes, probe and readout wave, respectively. The multiple parasitic peaks around pump and probe are higher-order harmonics coming from the electro-optic modulator. The frequency difference between pump and Stokes is set to 11.066 GHz − the Brillouin peak frequency of the [0,+1] mode. The generation efficiency of the readout reaches its maximum when the probe frequency is 488 MHz (with a theoretical value of 466 MHz) higher than that of the pump wave. The readout frequency, due to the phase matching condition, is also 488 MHz lower than Stokes wave. Fig. 3(b) shows the spectrum when pump/Stokes and probe/readout modes are exchanged. Unlike the previous configuration, in this case the probe frequency will be upshifted from the pump frequency; and since the dispersion curves of the two circularly polarized states are not parallel, the theoretical frequency separation should be less than 488 MHz. However, due to the weak circular birefringence, such tiny frequency deviation could not be resolved experimentally and the probe frequency shift $\Delta\Omega$ for maximum readout efficiency is still measured to be 488 MHz. Fig. 3(c) shows the readout power as a function of frequency difference between pump and Stokes when probe frequency shift $\Delta\Omega$ is set to 488 MHz. The result is in good agreement with the Lorentzian fitting, with a full width half maximum (FWHM) of 52 MHz, which is similar to the original Brillouin gain bandwidth of [0,±1] modes (55 MHz) in the fiber [18].

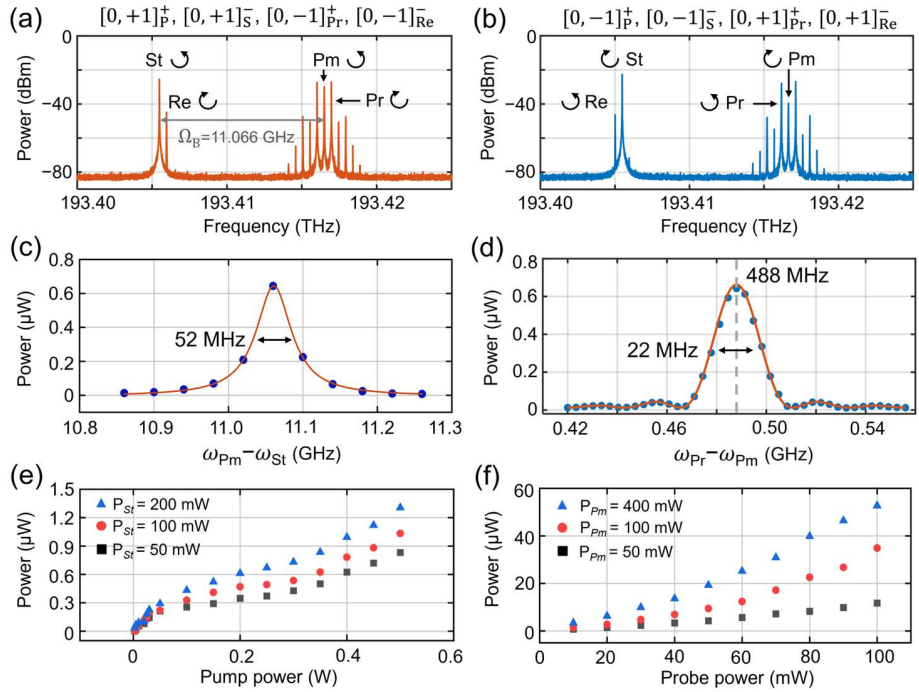

**Fig.** 3. (a) The measured spectrum (after the filter) when pump and Stokes modes are [0,+1]; (b) The measured spectrum when except pump and Stokes modes are [0,−1]. The superscripts "+" and "−" denote forward and backward propagation direction. The subscripts "Pm", "St", "Pr" and "Re" denote pump, Stokes, probe and readout waves; (c) The readout power as a function of pump-Stokes frequency difference; (d) The readout power as a function of probe-pump frequency difference; (e) Readout power changing with pump power, when Stokes power is 50 mW, 100 mW or 200 mW and probe power is 20 mW; (f) Readout power changing with probe power, when pump power is 50 mW, 100 mW or 400 mW and Stokes power is 100 mW.

Fig. 3(d) shows the readout power varying with probe-pump frequency difference $\Delta\Omega$ when pump-Stokes frequency difference is fixed at 11.066 GHz. A theoretical fitting to the sinc function shows that the FWHM bandwidth of the readout process is 22 MHz. Fig. 3(e) shows the readout power as a function of pump power, when the Stokes power is set to 50 mW, 100 mW and 200 mW and the probe power is set to 20 mW. The readout power first changes linearly with pump power in the low power regime, as the BDG produced at this point is very weak, and the readout process can be approximated as a conventional fiber Bragg grating diffraction process [22]. However, in the high pump power regime, the readout starts to vary nonlinearly (exponential growth) with pump power, because at this point, the BDG is sufficiently strong that it can amplify the readout wave through the secondary SBS process. Actually, as we already mentioned in the introduction, the writing and reading of BDG are four-wave mixing nonlinear processes, in which the four waves are always coupled together. Fig. 3(f) shows the readout power as a function of probe power, when pump power is 50 mW, 100 mW and 400 mW and stokes power is 100 mW. Again, it can be clearly seen that, the readout power starts to change in a nonlinear way with increasing probe when the pump power becomes stronger.

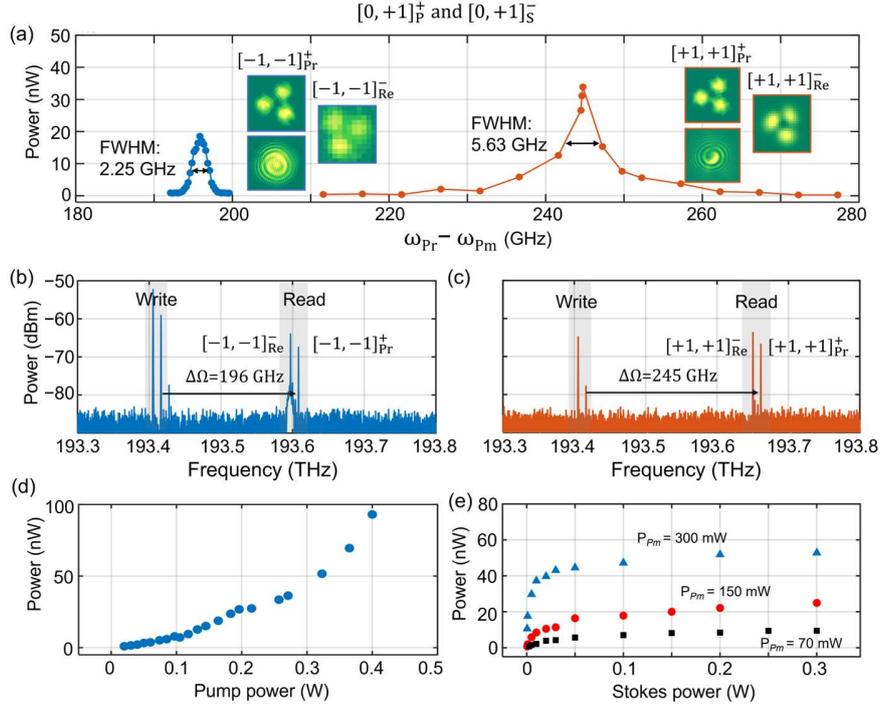

**Fig. 4.** (a) The power of $[-1,-1]^-$ and $[+1,+1]^-$ vortex-carrying readout signal as function of probe-to-pump frequency difference $\Delta\Omega$, when the pump and Stokes are fundamental modes $[0,-1]^+$ and $[0,-1]^-$. The insets show the measured mode profiles of two pairs of probe and readout signal. The probe signal is measured by a CCD camera and the readout signal is measured by the NBA system; (b) The measured spectrum showing BE-FWM when probe and readout are $[-1,-1]^+$ and $[-1,-1]^-$; (c) The measured spectrum when probe and readout are $[+1,+1]^+$ and $[+1,+1]^-$; (d) The readout power changing with pump power, when Stokes and probe power are 50 mW and 20 mW; (e) Readout power changing with Stokes power, when probe power is fixed at 20 mW and the pump power is 70 mW, 150 mW or 300 mW.

Next, we spatially modulate the probe wave with circularly-polarized vortex modes and study the intervortex BE-FWM in twisted PCF. The pump and Stokes modes are still $[0,-1]^+$ and $[0,-1]^-$ fundamental modes. Fig. 4(a) shows the measured power of readout wave as functions of probe-pump frequency difference ($\Delta\Omega$), when the probe waves are $[-1,-1]^+$ and $[+1,+1]^+$ vortex modes (from left to right). The insets in the figure show the mode profiles of $[-1,-1]^+$ probe and $[-1,-1]^-$ readout that were recorded during the measurement of left curve, and mode profiles of $[+1,+1]^+$ probe and $[+1,+1]^-$ readout recorded while measuring the right curve. The interference between probe modes and a divergent Gaussian mode are also shown in the insets, showing different vortex topology. The readout mode profiles were specifically measured by near-field scanning Brillouin analyzer (NBA) [18]. The maximum readout efficiencies for $[-1,-1]^+$ and $[+1,+1]^+$ vortex probes occur at $\Delta\Omega$=196 GHz and 245 GHz, corresponding to wavelength shifts of −1.58 nm and −1.97 nm. The experimentally measured frequency shifts match well with theoretical value of 190 GHz and 261 GHz. The FWHM bandwidth for $[-1,-1]^-$ and $[+1,+1]^-$ readout modes generations are 2.25 GHz and 5.63 GHz. Fig. 4(b) and 4(c) show the measured spectra after the output was reflected by PBS and purified by narrowband filter. The probe frequencies were set above the pump frequency of 196 GHz and 245 GHz. The left peaks in each figure are residual pump and Stokes during the BDG writing process with fundamental modes and the right peaks are vortex readout and residual probe signal in the BDG reading process. The tiny bump at the readout frequency in the Fig. 4(b) comes from parasitic Kerr effect in twisted PCF, which was not completely filtered out by the filter of 6 GHz transmission bandwidth. Fig. 4(d) shows the readout power as a function of pump power, when the Stokes power is 50 mW and probe power is 20 mW. Fig. 4(e) shows the readout power as a function of Stokes power, when the probe power is fixed at 20 mW and the pump power is 70 mW, 150 mW or 300 mW. It can be clearly observed that the readout efficiency gradually saturates as the Stokes power becomes larger. This is due to the fact that the strong Stokes wave depleted the pump so that the SBS effect has no more energy support and reaches equilibrium state. Nevertheless, the readout power can be further improved by increase the pump power (Fig. 4(e)).

Moreover, we also spatially-modulate the pump and Stokes with $[+1,+1]^+$ and $[+1,+1]^-$, and re-measure the readout power over probe-to-pump frequency difference $\Delta\Omega$ (Fig. 5(a)) and the spectrum (Fig. 5(b)). The measured frequency shift of the reading process is 64.5 GHz, showing again good agreement with the theoretical value of 70.2 GHz. In the experiment, the readout power somehow fluctuated due to the intermodal crosstalk, which makes the measured readout mode profile noisy.

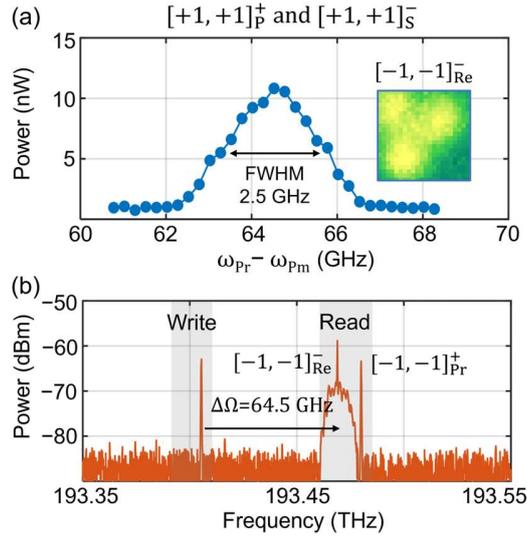

**Fig. 5.** (a) The power of $[-1,-1]^-$ vortex-carrying readout signal as a function of probe-to-pump frequency difference $\Delta\Omega$, when the pump and Stokes are vortex modes of opposite topological charges $[+1,+1]^+$ and $[+1,+1]^-$. The insets show the mode profile of readout mode measured by NBA system. (b) The measured spectrum showing BE-FWM process among $[+1,+1]^+$ pump, $[+1,+1]^-$ Stokes, $[-1,-1]^+$ probe and and $[-1,-1]^-$ readout. The tiny bump around the readout frequency comes from parasitic Kerr effect that was not fully filtered out. The pump signal has been completely filtered out, with only residual Stokes signal left in the spectrum.

Finally, we demonstrate a cross-frequency information-selective dynamic transfer from pump to readout under the mediation of BDG, by leveraging the strong circular polarization maintaining capacity of twisted PCF. The setup and its description are shown in the Appendix B. A series of binary information were loaded in the form of pulses onto the pump. Every five binary numbers correspond to a letter, with the code strategy shown in the Table1 in Appendix C, and the successive temporal segments covering each five binary numbers were multiplexed and alternatively modulated by two orthogonal circular polarization states, as shown in Fig. 6. The information "1" has a pulse width of 0.5 μs and duty-cycle of 20% and "0" has no pulse.

The interaction between the information-encoded pump and CW Stokes will generate a pulsed-BDG carrying the same information, which will then be read by a CW probe, with the information being transferred to a readout signal at a different frequency than the pump frequency. When the CW probe carries right circular polarization states (RCP), the information encoded on LCP pump was selectively readout from BDG, showing "BilUN" according to the specific coding strategy from Table1. When the CW probe carries left circular polarization states (LCP), the information encoded on RCP pump was selectively readout, showing "rlOI" after decoding. Finally, when the CW probe is in the linear polarization (LP), the full information of the 'BrillOUIN' is read out. This cross-frequency dynamic information transfer with inter-circular polarization BDG shows again the robust circular polarization preserving ability and decent circular polarization-selective SBS in twisted PCF. The weak circular birefringence ($3.5\times10^{-6}$) in twisted PCF makes the group delays of two orthogonal circular polarization states negligible and enables the LCP and RCP pulses sequentially travel through the fiber without temporal overlap. The results can be potentially used in many areas, for example, frequency-shifted light storage, frequency conversion of quantum state as well optical communications.

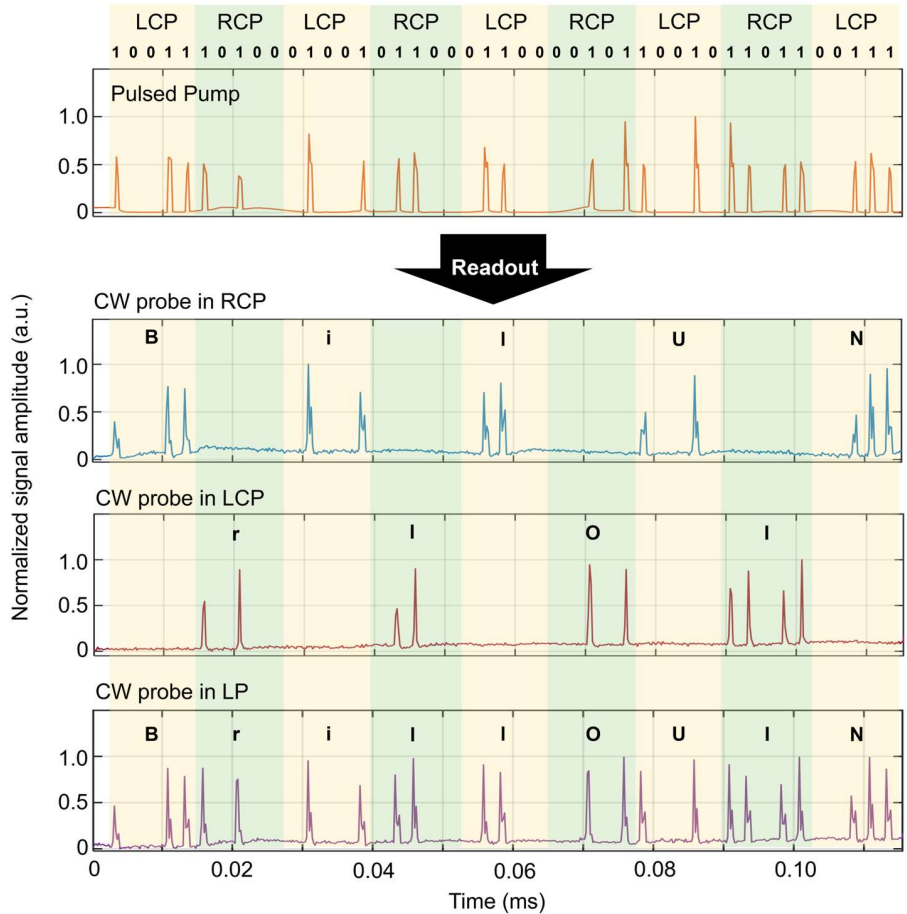

**Fig. 6.** The experiment results of dynamic information transfer from pump to frequency-shifted readout signal, with the mediation of BDG. The Stokes and probe are CW. The pump is modulated by a series of customized pulses that correspond to binary numbers, and two orthogonal circular polarization states are successively loaded on every five binary numbers, which denote one specific letter in "BrillOUIN" (Appendix C). Probe signals carrying different circular polarizations will selectively readout different information from the pump.

## 3. Conclusion

Inter-circular-polarization and intervortex BE-FWM have been demonstrated in twisted PCF, with the mediating BDG being excited by circular-polarization-selective and topology-selective SBS. The twisted PCF can stably transmit chiral optical states: SAM (i.e. circular polarization states) and OAM (i.e. vortex states), enabling the observation of chiral BE-FWM effect. The BDG generated by SBS between two counterpropagating circular polarization/vortex states can diffract a probe wave, whose frequency as well as the circular polarization state/vortex state are different from the pump. In addition, dynamic cross-frequency information transfer from pump to readout using the inter-circular-polarization BE-FWM effect has been demonstrated, showing excellent information-selective readout capability and robust circular polarization preservation ability.

When the pump and Stokes modes are fixed, the readout efficiency varies considerably depending on the probe mode. This is due to the different optoacoustic overlap between the probe mode and the travelling acoustic grating. Apart from that, we note that the linear polarization maintaining fiber (PMF) may also be used to realize the cross-frequency information transfer, although due to the strong linear birefringence of such fibers ($>1\times10^{-4}$), one may need to greatly increase the time interval between two pulses to avoid information overlap. Finally, the results of BE-FWM with chiral optical states open a new perspective for BE-FWM and might be applicable in the fields of quantum information processing and optical communications.

**Acknowledgements.** The authors acknowledge the funding from Max-Planck-Gesellschaft through an independent Max-Planck-Research Group. The authors thank Michael H. Frosz for providing the twisted PCF and Philip St.J. Russell for helpful discussions.

**Conflict of Interest.** The authors declare no conflicts of interest.

**Data Availability.** Data underlying the results presented in this paper are not publicly available at this time but may be obtained from the authors upon reasonable request.

**Author Contributions.** X.Z. and B.S. proposed the concept and designed the experimental plan. X.Z. built the experimental setup and performed the measurements. X.Z. and B.S. preformed the data analysis and wrote the manuscript. B.S. led the project.

**Appendix A: Refractive indices of all six chiral modes**

The calculated refractive indices of all six modes are plot in the Fig. 7(a) and the zoom-in plots for each mode are shown in Fig. 7(b), (c) and (d). The [0,+1] and [0,−1] at 1550 nm are 1.4415323 and 1.4415288, and the refractive indices of [−1,+1], [−1,−1], [+1,+1] and [+1,−1] at 1550 nm are 1.4401163, 1.4401112, 1.4395986, 1.4395935.

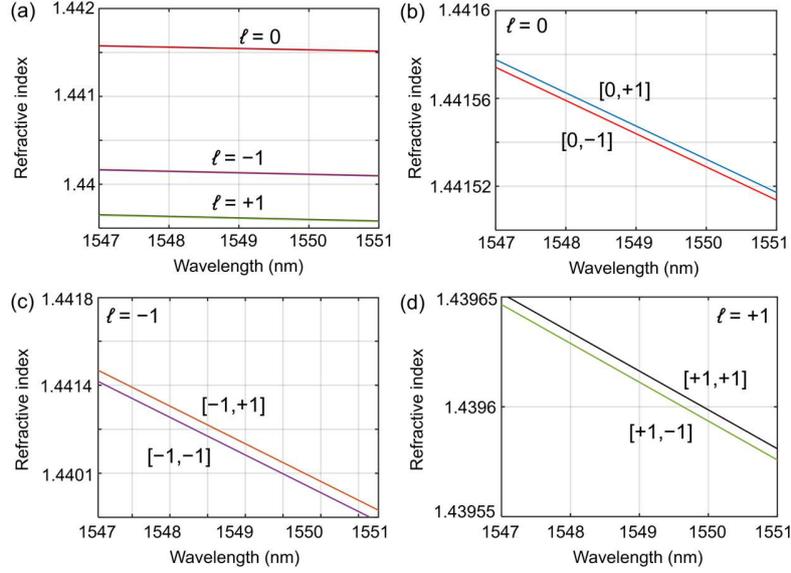

**Fig. 7.** (a) Numerically calculated mode refractive indices of $\ell=0$, $\ell=+1$ and $\ell=-1$ modes. The first one has two circularly-polarized fundamental modes and the latter two has circularly-polarized vortex modes; (b) The zoom-in plot for $\ell=0$ modes; (c) The zoom-in plot for $\ell=+1$ modes; (d) The zoom-in plot for $\ell=-1$ modes.

**Appendix B: Experimental setup of cross-frequency selective information transfer**

The experimental setup is shown in the Fig. 8(a). The Stokes and probe signals are CW light and the generation of them are the same as the Fig. 2(a). The pump is split into two paths, which are pulsed modulated by an arbitrary wavefunction generator (AWG). Each path has pulse coding and non-coding regions, and when the two paths are combined with a beam splitter, the non-coding region of one path is filled with the coding region of the other path, as shown in Fig. 8(b). In order to make sure the coding regions of both paths do not overlap after beam combination, both pulse trains are synchronized and their time delays are fine-tuned by AWG. Path 1 is loaded with left circular polarization state and path 2 with right circular polarization state. The second beam splitter is used to combine the pump with a probe, the polarization of which is tuned to selectively read out the information in the pump. After all other frequencies have been filter out by a fiber Bragg grating (FBG) of 210 MHz FWHM transmission bandwidth, the readout wave is finally collected by a photodetector and oscilloscope.

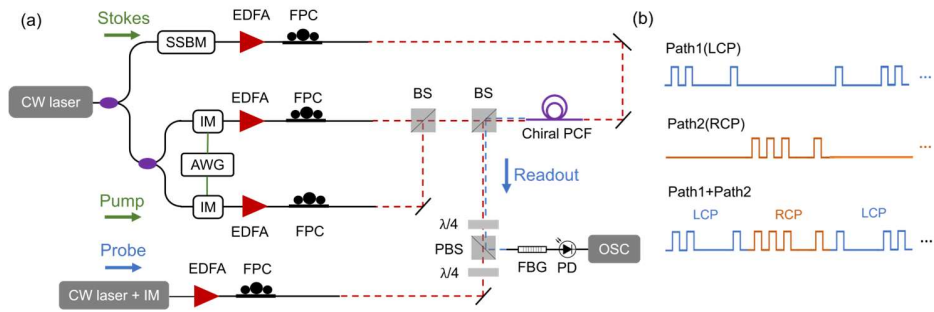

**Fig. 8.** (a) The experiment setup of cross-frequency selective information transfer with inter-circular polarization BE-FWM. (b) Schematic illustration of combination of two pump pulse trains with orthogonal circular polarization states.

**Appendix C: Coding strategy for letters "BrillOUIN"**

| Letter | Code |
|---|---|
| B | 10011 |
| r | 10100 |
| i | 01001 |
| l | 01100 |
| O | 00101 |
| U | 10010 |
| I | 11011 |
| N | 00111 |